\documentclass[paper=a4, 11pt, twoside, twocolumn]{scrartcl}
\usepackage[utf8]{inputenc}
\usepackage[english]{babel}
\usepackage{textcomp}

\usepackage[left=3cm, right=2cm, top=2cm, bottom=3cm]{geometry}

\usepackage{graphicx}
\usepackage{subcaption}

\usepackage{tikz}
\usetikzlibrary{positioning, shapes}

\usepackage[backend=bibtex, sorting=none, style=numeric-comp,maxbibnames=1,firstinits=true,url=false]{biblatex} 
\addto\captionsenglish{}

\usepackage{authblk}

\title{Experimental realization of type-II Weyl state in non-centrosymmetric TaIrTe$_4$}
\author[1]{E. Haubold}
\author[1]{K. Koepernik}
\author[1]{D. Efremov}
\author[1,2]{S. Khim}
\author[1,3]{A. Fedorov}
\author[1]{Y. Kushnirenko}
\author[1,4]{J. van den Brink}
\author[1,4]{S. Wurmehl}
\author[1,4]{B. Büchner}
\author[5]{T. K. Kim}
\author[5]{M. Hoesch}
\author[6]{K. Sumida}
\author[6]{K. Taguchi}
\author[6]{T. Yoshikawa}
\author[6]{A. Kimura}
\author[7]{T. Okuda}
\author[1]{S. V. Borisenko}

\affil[1] {IFW Dresden, P.O. Box 270116, 01171 Dresden, Germany}
\affil[2] {Max-Planck-Institut für Chemische Physik fester Stoffe, Nöthnitzer Straße 40, 01187 Dresden, Germany}
\affil[3] {II. Physikalisches Institut, Universität zu Köln, Zülpicher Straße 77, 50937 Köln, Germany}
\affil[4] {Department of Physics, TU Dresden, 01062 Dresden, Germany}
\affil[5] {Diamond Light Source, Harwell Campus, Didcot OX11 0DE, United Kingdom}
\affil[6] {Graduate School of Science, Hiroshima University, 1-3-1 Kagamiyama, Higashi-Hiroshima 739-8526, Japan}
\affil[7] {Hiroshima Synchrotron Radiation Center (HSRC), Hiroshima University, 2-313 Kagamiyama, Higashi-Hiroshima 739-0046, Japan}

\date{\vspace{-5ex}}

\begin{document}
\twocolumn[\begin{@twocolumnfalse}
	\maketitle
	\begin{abstract}
		Recent breakthrough in search for the analogs of fundamental particles in condensed matter systems lead to experimental realizations of 3D Dirac and Weyl semimetals \cites{svbprl,neupane, dingTaAs, hasanTaAs, chenTaAs}{borisenko2016yb}. Weyl state can be hosted either by non-centrosymmetric or magnetic materials and can be of the first or the second type \cite{wan2011, weng2015, balents2011, soluyanov2015type}. Several non-centrosymmetric materials have been proposed to be type-II Weyl semimetals, but in all of them the Fermi arcs between projections of multiple Weyl points either have not been observed directly or they were hardly distinguishable from the trivial surface states which significantly hinders the practical application of these materials \cite{xu1603, jiang1604, liang1604, deng1603, huang1603, xu1604, bruno1604, wang1604, wu1604, feng1606, tamai2016}. Here we present experimental evidence for type-II non-centrosymmetric Weyl state in TaIrTe$_4$ where it has been predicted theoretically \cite{koepernik2016tairte}. We find direct correspondence between ARPES spectra and calculated electronic structure both in the bulk and the surface and clearly observe the exotic surface states which support the quasi-1D Fermi arcs connecting only four Weyl points. Remarkably, these electronic states are spin-polarized in the direction along the arcs, thus highlighting TaIrTe$_4$ as a novel material with promising application potential. 
	\end{abstract}           
	\vspace{0.5cm}
\end{@twocolumnfalse}
]

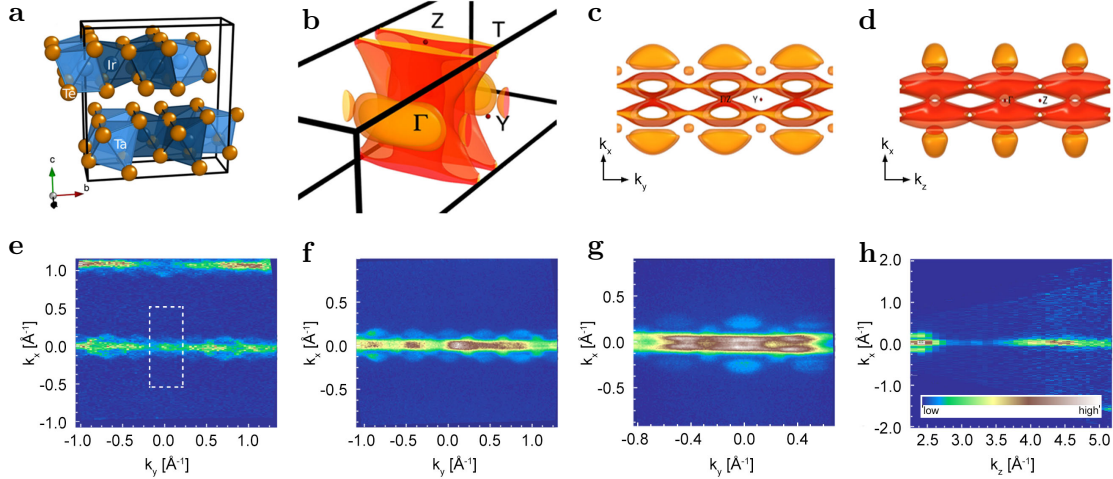
\begin{figure*}[!ht]
	\centering
	\begin{tikzpicture} [bla/.style={%
		text height = 9.5em,
		text depth = .25em,
		text centered
	}]
		\node[] (na) at(0,0) {
			\includegraphics[width=.9\textwidth]{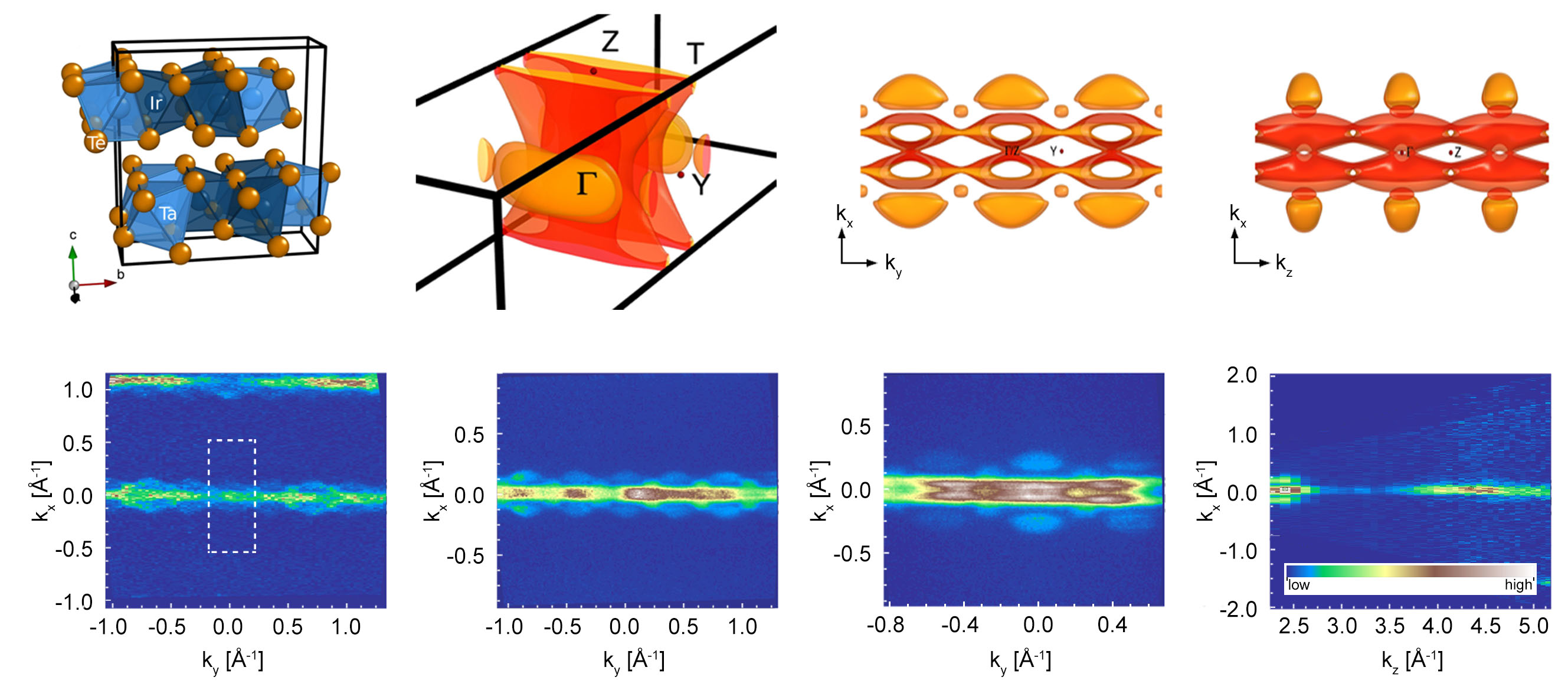}
		};

		\node[ anchor= north west] (la) at(-7.5,3.2) {\bf{a}};
		\node[ anchor= north west] (lb) at(-3.6,3.2) {\bf{b}};
		\node[ anchor= north west] (lc) at(0.2,3.2) {\bf{c}};
		\node[ anchor= north west] (ld) at(3.8,3.2) {\bf{d}};
		\node[ anchor= north west] (le) at(-7.5,0.1) {\bf{e}};
		\node[ anchor= north west] (lf) at(-3.6,0.1) {\bf{f}};
		\node[ anchor= north west] (lg) at(0.2,0.1) {\bf{g}};
		\node[ anchor= north west] (lh) at(3.8,0.1) {\bf{h}};
	\end{tikzpicture}
	\caption{(a) Crystal structure of TaIrTe$_4$. (b) Calculated 3D Fermi surface, (c) Projection to $k_x$-$k_y$ plane, (d) Projection to $k_x$-$k_z$ plane. ARPES Fermi surface maps taken using different photon energies: (e) 100 eV with Brillouin-zone shown as dashed box, (f) 70 eV and (g) 25 eV. (h) Photon energy dependent ARPES scan from 20eV to 100eV. A colorscale used in panels (e)-(h) is shown as an inset.} 
	
	\label{fig:one}
\end{figure*} 

Weyl semimetals are compounds hosting Weyl fermions, which have a fundamental importance. These fermions are quasiparticles described by the Weyl equation \cite{weyl1929} thus being a condensed-matter realization of the elusive Weyl particles, long sought in high-energy physics. The exotic behavior is attributed to the electronic states emerging due to crossings of non-degenerate bands in a single point of k-space, the Weyl point, which can be realized in the materials where either time-reversal or inversion symmetry is broken \cite{wan2011, weng2015, balents2011}. The simplest Fermi surface of such a Weyl semimetal would consist of only two such points \cite{wan2011}.


Recently, a new type of Weyl semimetals has been introduced and experimentally detected \cite{soluyanov2015type,borisenko2016yb}. Unlike in the standard type-I materials, the Weyl cone is strongly tilted, such that the Fermi surface consists now of both electron-like and hole-like pockets touching in a single point, the Weyl point, which remains at the Fermi level. This qualitative distinction leads to marked differences in the thermodynamics and response to magnetic fields. For example, the chiral anomaly \cite{nielsen1983} will be observed only if the direction of the magnetic field is within the cone made by touching Fermi surfaces \cite{soluyanov2015type}. Moreover, the topological surface states connecting the Weyl points, the Fermi arcs, should have a spin texture which could lead to exotic surface transport and contribute to extremely large magnetoresistance \cite{jiang1503}.
Thus, apart from the fundamental significance of type-II Weyl state, the experimental realization of a material allowing to control the direction of the conducted current with a magnetic field would be an important gain for future applications. 

It has been recently predicted \cite{koepernik2016tairte} that TaIrTe$_4$ is a ternary Weyl semimetal of type II. The crystal structure of this material is shown in Fig. \ref{fig:one}(a). It is a non-centrosymmetric stacked crystal with covalent bonds along axes a and b and Van-der-Waals bonds along axis c, making it very suitable for experimental techniques which require atomically clean surfaces. Extensive band-structure calculations have been performed for the bulk and two surfaces ($(001)$ and $(00\bar{1})$ cleavage planes) of this material \cite{koepernik2016tairte}.  The Fermi surface shown in Fig. \ref{fig:one}(b-d) consists of several sheets: three-dimensional and well isolated from other hole-like pockets (light orange colors) and quasi-2D electron-like pockets (dark orange to red colors). It is the boundaries of the large 3D hole-like pockets, where the calculations have identified two pairs of Weyl points, each connected by long Fermi arcs \cite{koepernik2016tairte}. In the following we will demonstrate the presence of these arcs experimentally. 

\begin{figure*}[!ht]
	\centering
	\begin{tikzpicture}
		\node[] (na) at(0,0) {
			\includegraphics[width=.9\textwidth]{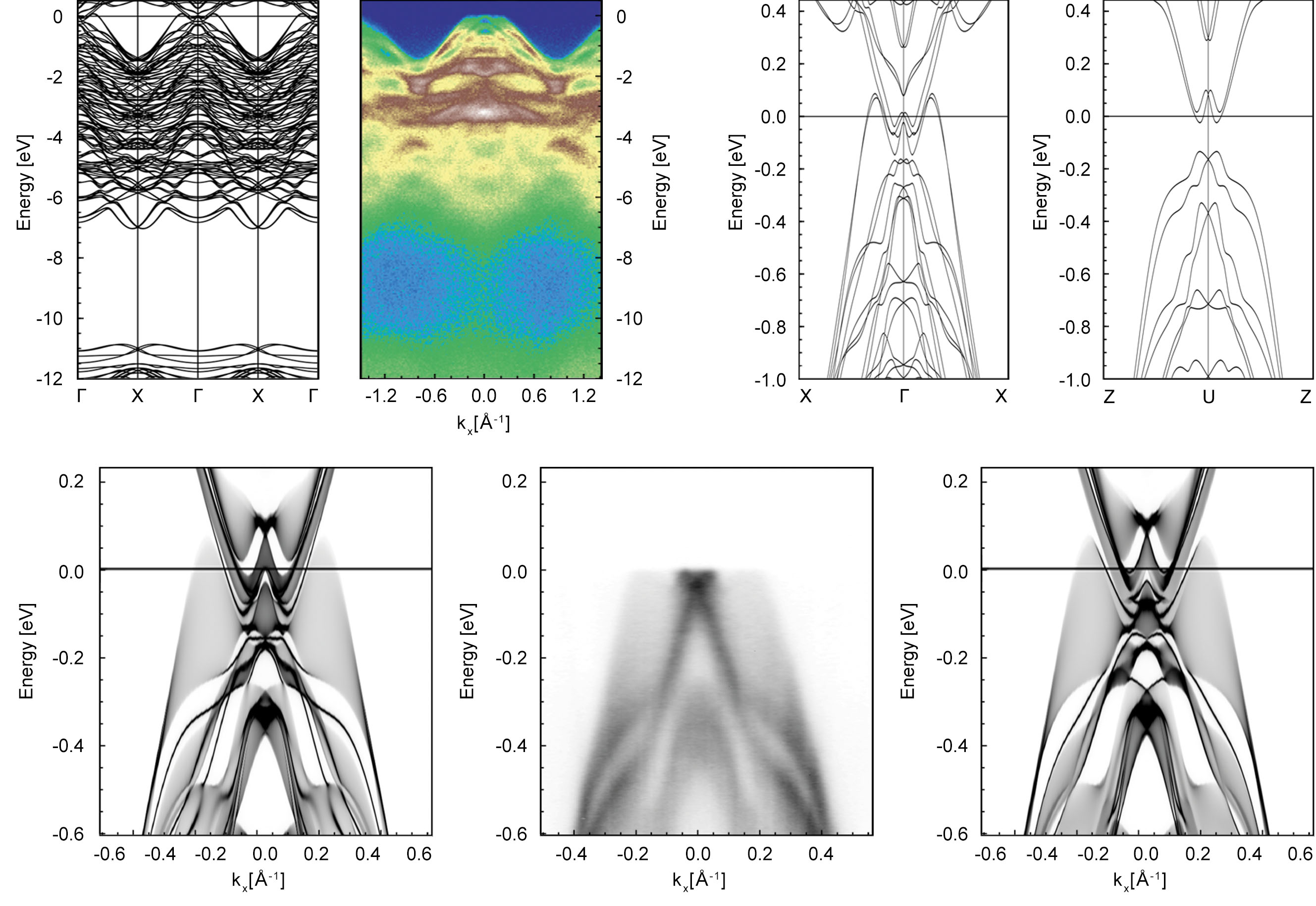}
		};

		\node[anchor= south west] (la) at(-7.5,4.7) {\bf{a}};
		\node[anchor= south west] (lb) at(-3.9,4.7) {\bf{b}};
		\node[anchor= south west] (lc) at(0.3,4.7) {\bf{c}};
		\node[anchor= south west] (ld) at(4.1,4.7) {\bf{d}};
		\node[anchor= south west] (le) at(-7.4,-0.6) {\bf{e}};
		\node[anchor= south west] (lf) at(-2.3,-0.6) {\bf{f}};
		\node[anchor= south west] (lg) at(2.7,-0.6) {\bf{g}};
	\end{tikzpicture}
	\caption{(a) Band structure calculations along X-$\Gamma$-X. (b) ARPES measurements along X-$\Gamma$-X. (c) Zoomed-in band structure along  X-$\Gamma$-X, (d) Zoomed-in band structure along Z-U-Z. (e) Calculated spectral function at $k_y=0$ for the $(001)$ surface. (f) Corresponding ARPES data taken at $99eV$ photon energy. (g) Calculated spectral function at $k_y=0$ for the $(00\bar{1})$ surface. }
	\label{fig:two}
\end{figure*}
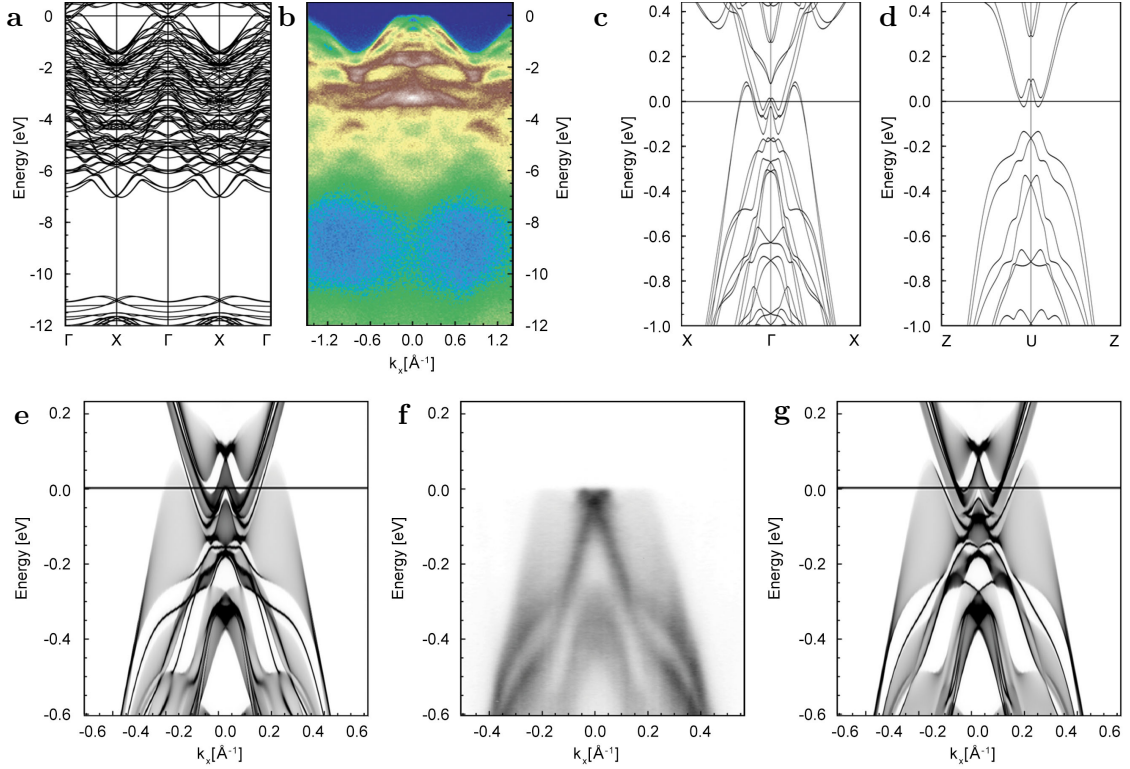

TaIrTe$_4$ single crystals were grown from excess Te flux. Ta, Ir, and Te were mixed and placed in a crucible which was sealed inside a quartz tube under vacuum. The tube was heated up to $1000$ \textdegree C, slowly cooled down to $700$\textdegree C, and then centrifuged to separate excess Te from the grown crystals \cite{khim2016}.
In order to have a complete comparison between the theory and experiment in three dimensions, we have studied the single crystals of TaIrTe$_4$ by angle-resolved photoemission spectroscopy (ARPES) using synchrotron light with variable photon energies. Normal ARPES measurements were performed at beamline I05 of the Diamond Light Source using a Scienta R4000 hemispherical electron enery analyzer with an angular resulution of $0.20$\textdegree\ and $3$ meV in energy resolution. The samples were cleaved in situ in a vacuum lower than $2\times 10^{-10}$ mbar and were measured at temperatures between $5$ to $10$K.
Spin resolved data was recorded at the ESPRESSO Spin ARPES detector at Hiroshima Synchrotron Radiation Center. The system uses a scienta R4000 electron energy analyzer and has two VLEED detectors connected to it, each able to measure Spin polarization in two axes. Overall the system is able to measure a complete 3D spin resolved image of the material \cite{espresso, okuda2015}.
%
The band structure calculations were obtained using the Full Potential Local Orbital code (FPLO) \cite{koe99} using the setup from \cite{koepernik2016tairte}. 
 From the DFT calculation a minimum basis Wannier function model was extracted, which then was mapped onto a semi-infinite slab geometry whose surface spectral function was obtained via Greens function techniques.

In Fig. \ref{fig:one}(e-h) we show the ARPES Fermi surface maps measured from the cleaved $(001)$ surface. The overview map recorded using $100eV$ photons is shown in Fig. \ref{fig:one}(e) together with the Brillouin zone (BZ). As it is seen from the map, the electronic structure naturally follows the significant anisotropy of the crystal structure and is in a remarkable agreement with the results of the calculations. The Fermi surface is relatively small, implying rather low concentration of charge carriers. Large, non-symmetrized maps with subsequent zooming-in (panels e-g) by means of lowering the photon energy allowed us to effectively scan the electronic structure at different $k_z$ values. While there are no big qualitative changes, the size, shape and intensity of particular features are not constant. For example, the isolated areas almost homogeneously filled by intensity and clearly corresponding to hole-like pockets (light-orange surfaces in Fig. \ref{fig:one}(b-d)) appear differently in different maps and in neighboring BZs within the same map and in some cases are hardly observed at all.

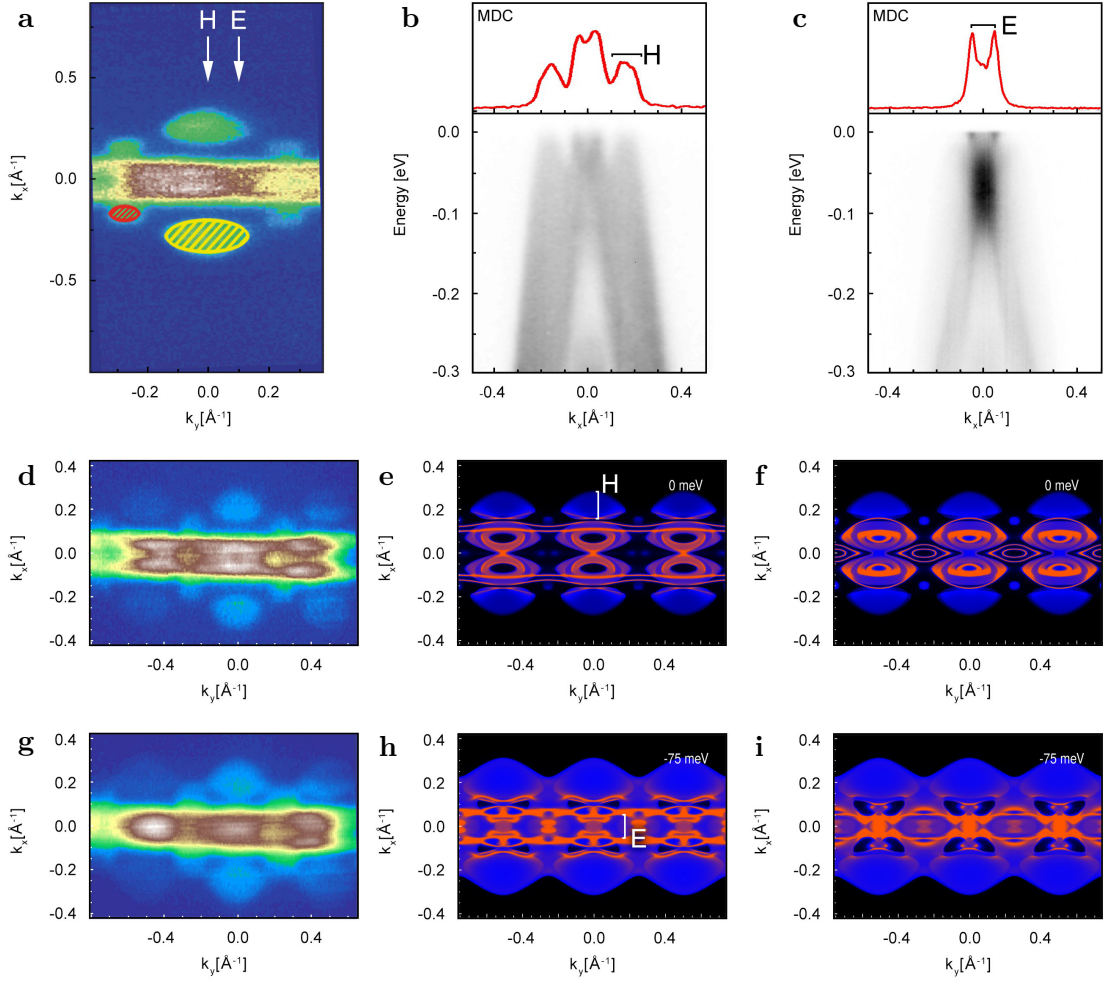
\begin{figure*}[!ht]
	\centering
	\begin{tikzpicture}
		\node[] (na) at(0,0) {
			\includegraphics[width=.9\textwidth]{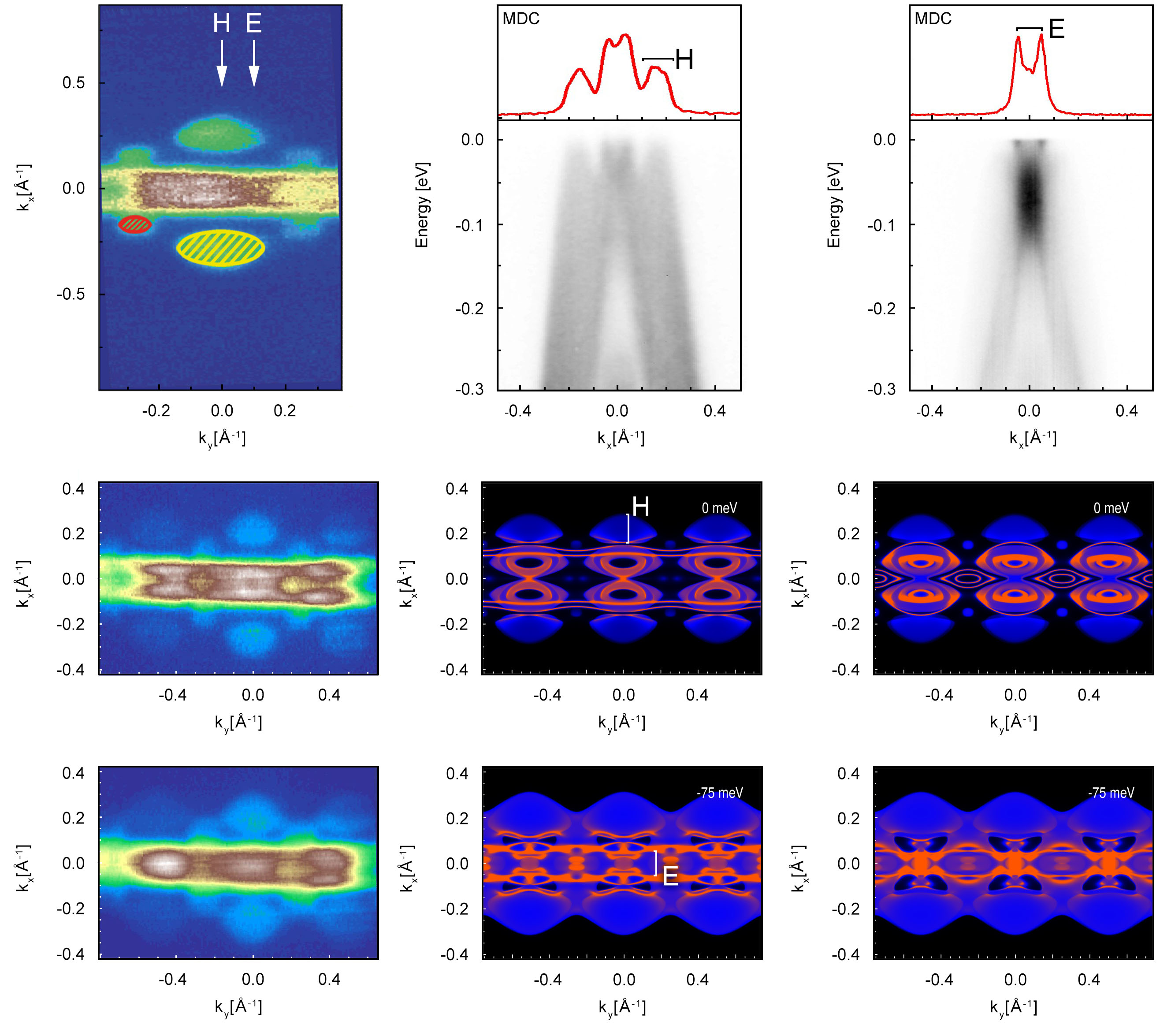}
		};
	
		\node[anchor= south west] (la) at(-7.3,6.1) {\bf{a}};
		\node[anchor= south west] (lb) at(-2.2,6.1) {\bf{b}};
		\node[anchor= south west] (lc) at(3,6.1) {\bf{c}};
		\node[anchor= south west] (le) at(-7.3,0) {\bf{d}};
		\node[anchor= south west] (lf) at(-2.5,0) {\bf{e}};
		\node[anchor= south west] (lg) at(2.5,0) {\bf{f}};
		\node[anchor= south west] (li) at(-7.3,-3.6) {\bf{g}};
		\node[anchor= south west] (lj) at(-2.5,-3.6) {\bf{h}};
		\node[anchor= south west] (lk) at(2.5,-3.6) {\bf{i}};
	\end{tikzpicture}
	\caption{(a) ARPES Fermi surface map with the shaded areas of hole pockets for comparison with dHvA. The Arrows indicate the positions of the cuts shown in b (H) and c (E). (b) and (c) ARPES momentum-energy cuts at positions shown by H and E in panel (a) with corresponding Fermi-level MDC curves on top. (d) ARPES Fermi surface map taken using 25 eV photons and covering three BZs. (e) Calculated Fermi surface map for the $(001)$ surface. (f) Calculated Fermi surface map for the $(00\bar{1})$ surface. (g) ARPES Momentum distribution at $25meV$ below Fermi level. (h) Calculated momentum distribution of the intensity at $75meV$ binding energy for the $(001)$ surface. (i) Calculated momentum distribution of the intensity at $75meV$ binding energy for the $(00\bar{1})$ surface.}
	\label{fig:three}
\end{figure*}

This is in agreement with their three-dimensional character predicted by the calculations - the ARPES momentum resolution along $k_z$ does not allow one to observe sharp contours as in the case of quasi-2D features, but its finite value still makes it possible to track the mentioned changes. We have also recorded a detailed 
excitation energy dependence map of the center of the Brillouin-zone around the $\Gamma$-point. The resulting $k_x$-$k_z$ projection of the Fermi surface is shown in Fig. \ref{fig:one}(h). There is a periodically changing intensity pattern which roughly corresponds to the calculated projection from Fig. \ref{fig:one}(d). Quasi-two-dimensional electron-like pockets are clearly seen in all maps as well, more or less precisely mimicking the repeating arrangement from the upper panels of Fig. \ref{fig:one}. Thus, experimental Fermi surface qualitatively agrees with the calculated one.


The next step is to estimate the degree of quantitative agreement between the calculations and the measured data. We present momentum-energy intensity maps which show the underlying dispersions in Fig. \ref{fig:two}. The comparison of the datasets corresponding to X-$\Gamma$-X cut on the large energy scale is shown in Fig. \ref{fig:two}(a, b). The good agreement is emphasized by correctly captured by the calculations bandwidth ($\sim$ 7 eV) and overall pattern of the numerous dispersions. Comparison on the smaller scale of hundreds of meV involves two ARPES-specific effects which need to be taken into account. First is the mentioned above three-dimensionality of the TaIrTe$_4$ and moderate $k_z$-resolution of the method. The degree of $k_z$-dispersion is seen from the band structure along X-$\Gamma$-X and the Z-U-Z cuts (Fig. \ref{fig:two}(c, d)). Second is the presence of the surface which may result in occurrence of the surface states. Our calculations of the spectral function of an ideal semi-infinite slab consider both factors. In Fig. \ref{fig:two}(e-g) we show the photoemission data along the cut through the center of the BZ together with the spectral function corresponding to the two opposite surfaces. Although the overall agreement is striking, the closer inspection reveals the differences in the vicinity of the Fermi level: the crossing of the linear dispersions in the experiment occurs at lower binding energies and the corresponding fine structure is not present/resolved. This observation implies that still higher resolution data are needed to figure out, e. g., which surface is probed and whether the surface states are present.

\begin{figure*}[!ht]
	\centering
	\begin{tikzpicture}
		\node[] (na) at(0,0) {
			\includegraphics[width=.9\textwidth]{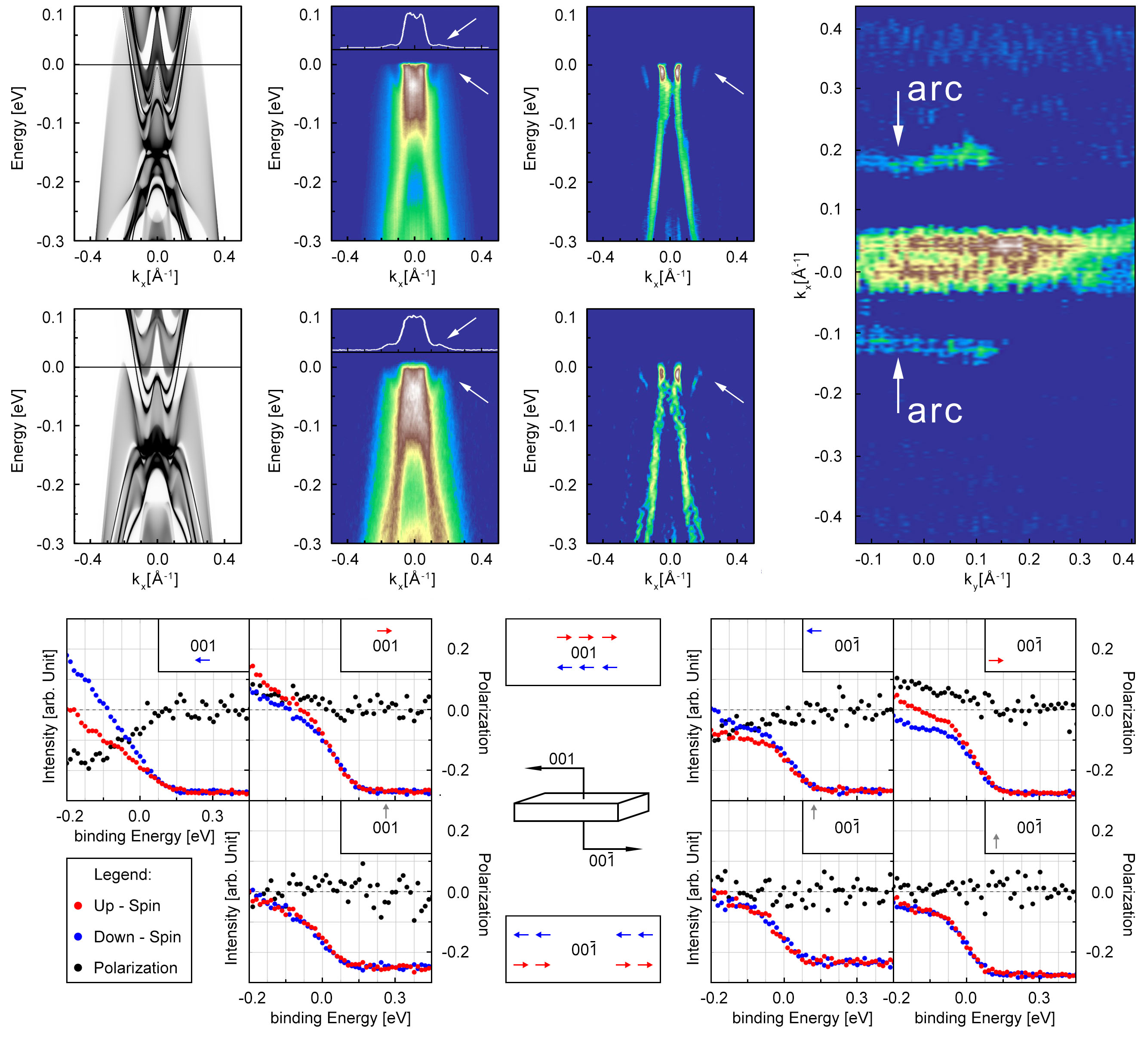}
		};
		\node[anchor= south west] (la) at(-7.55,6.5) {\bf{a}};
		\node[anchor= south west] (lb) at(-4.2,6.5) {\bf{b}};
		\node[anchor= south west] (lc) at(-0.8,6.5) {\bf{c}};
		\node[anchor= south west] (ld) at(-7.55,2.5) {\bf{d}};
		\node[anchor= south west] (le) at(-4.2,2.5) {\bf{e}};
		\node[anchor= south west] (lf) at(-0.8,2.5) {\bf{f}};
		\node[anchor= south west] (lg) at(2.8,6.5) {\bf{g}}; 
		\node[anchor= south west] (lh) at(-6.7,-1.6) {\bf{h}};
		\node[anchor= south west] (li) at(-4.2,-1.6) {\bf{i}};
		\node[anchor= south west] (lj) at(-4.2,-4.1) {\bf{j}};
		\node[anchor= south west] (ll) at(1.8,-1.6) {\bf{l}};
		\node[anchor= south west] (lm) at(4.2,-1.6) {\bf{m}};
		\node[anchor= south west] (ln) at(1.8,-4.1) {\bf{n}};
		\node[anchor= south west] (lo) at(4.2,-4.1) {\bf{o}};
		\node[anchor= south west] (lk) at(-1.4,-1.6) {\bf{k}};
	\end{tikzpicture}
	
	\caption{(a) Spectral function calculated for $k_y < k_y(WP)$ and (d) for $k_y > k_y(WP)$. (b, e) Corresponding ARPES data. Peaks in MDCs and dispersing features are shown by white arrows. (c, f) Second derivatives of the panels (b) and (e) highlighting the presence of the topological surface states. (g) Second derivative of the Fermi surface map of the second BZ, similar to the one shown in Fig. 3d. Arcs are clearly seen and are shown by the arrows. The same arrows point to the position of the arcs probed by SARPES. (h-j, l-o) Spin-resolved ARPES measurements performed at the points of the k-space schematically indicated by arrows next to the surface vector in the insets. (k) Schematically summary of the SARPES results. The electronic states on the arcs are spin-polarized as shown by the arrows within the BZ boundaries.}
	\label{fig:four}
\end{figure*}

To further quantify the discrepancies between the theory and experiment for the easier identification of the Weyl points and arcs, we analyzed the data taken with the lower photon energies which guarantee better energy and momentum resolutions. We compare more rigorously the calculated and experimental FSs in Fig.\ref{fig:three}. A high resolution Fermi surface map is shown in Fig. \ref{fig:three}(a) from which we can obtain the area covered by particular Fermi surface sheets. The areas of the small and large hole-like pockets are approximated by red and yellow ellipses and if calculated according to Ref. \cite{onsager1952}, are equal to $\sim$70 T and $\sim$215 T respectively. Both values are in a very reasonable agreement with the corresponding frequencies ($\sim$73 T and $\sim$240 T) determined in recent dHvA experiments \cite{khim2016}. As quantitative measures of the linear size of the large hole-like and electron-like FSs we consider the width of the momentum-distribution curve (MDC) introduced in Fig. \ref{fig:three}(b) and distance between the peaks of the MDC shown in Fig. \ref{fig:three} (c), respectively. Now we can directly compare the experimental FS (Fig. \ref{fig:three}(d)) with the Fermi surface map calculated from the spectral function for both surfaces (Fig. \ref{fig:three}(e, f)). From this comparison it is seen that while the size of the large hole-pockets (H) along $k_x$ is in a good correspondence with the predicted value, the small hole-like pockets appear noticeably smaller and electron-like ones (E) larger in the calculations. To match the size of the electron-like pockets one should cut the calculated spectral function at binding energy of 75 meV, as demonstrated in Fig. \ref{fig:three}(h, i). However, such a shift results in merging of all hole-like pockets, which are well separated in the experiment (Fig. \ref{fig:three}(a) and (d)). In turn, the merging hole-like pockets can be achieved by considering a momentum distribution of the ARPES intensity at 25 meV binding energy, as is seen in Fig. \ref{fig:three}(g). In other words, while all the features are present and look like those in the calculations, there is no single rigid shift which would result in a full quantitative agreement. In order to obtain the experimental electronic structure one needs to distort the theoretical one in a peculiar way: the electronic states making the small hole-like pockets should be shifted up by $\sim$25 meV and finally the dispersions responsible for the electron-like FS should be shifted up by 75 meV. We note, that such distortion of the calculated band structure is also routinely seen in iron-based superconductors \cite{borisenko2015direct} and in a certain sense is more natural than the rigid shift as it can preserve the charge.  From Fig. \ref{fig:three} we also learn that only the calculations for the $(001)$ surface (Fig. \ref{fig:three}(e) and (h)) are reproducing the experimental data, taking into account the aforementioned adjustment of different FS sheets. For example, the nearly parallel structures in the center of the Brillouin-zone are only seen from the $(001)$ surface maps and not in $(00\bar{1})$ ones (Fig. \ref{fig:three}(f) and (i)).


With the details of the electronic structure now at hand and degree of quantitative correspondence between theory and experiment being established, we can focus on the detection of the topological surface states which support Fermi arcs connecting the Weyl points. According to the calculated spectral function (see Fig. \ref{fig:three}(e) and Fig. 3 in Ref. \cite{koepernik2016tairte}), these surface states in the case of $(001)$ surface are located at the edge of the large hole-like FS. We do observe such edges at certain $k_z$ (e.g.: the right Brillouin-zone in Fig. \ref{fig:three}(d)) and in Fig. \ref{fig:four} we present more detailed and direct evidence for topological surface states supporting the Fermi arcs and the arcs themselves. %
In panels (a) and (d) of Fig. \ref{fig:four} we show theoretical calculated data for a cut through the fermi arc (a) and for a position with $k_y > k_y(WP)$, comparing them with two experimental cuts in momentum space which cross the FS arcs, shown in panels (b, c, e, f). 
One can easily identify the topological surface states as sharpening of the intensity on the inner sides of the smeared out intensity corresponding to 3D dispersions of the hole-pockets. This is confirmed by the maxima in MDCs ( white solid lines in panels (b) and (e) ) and by the second-derivative plots in panels (c) and (f).

The agreement with the calculations is both qualitative and quantitative, as the Fermi velocity is $\sim$ 1 eV\AA\ in all cases. 
In Fig. \ref{fig:four} (g) we show the second derivative of the FS map, which is similar to the right part of the one shown in Fig. 3d. Now the Fermi arcs are seen directly and with unprecedented clarity.

As is predicted theoretically and shown experimentally for $(001)$ surface, the arcs in TaIrTe$_4$ are peculiar: they are long and reasonably straight. With these characteristics in mind one would obviously be interested to learn about the spin of these states from the point of view of applications in spintronics.

In Fig. \ref{fig:four} (h-n) we present the spin-resolved measurements from both surfaces of TaIrTe$_4$. We have analyzed the energy distribution curves which correspond to the topological surface states (momentum integration window covers the single dispersion branch completely) in different parts of the BZ. There is always a clear signal corresponding to a particular direction of spin. The positive (negative) spin-polarization of a particular portion of the arc is given by red (blue) arrows while absence of spin-polarization is indicated by grey arrows. As it follows from Fig. \ref{fig:four} (h-j) for surface $(001)$ and from Fig. \ref{fig:four} (k-n) for surface $(00\bar{1})$, the spins are always directed along the arcs and their directions are naturally opposite when $k_y$ is inverted. We summarize our observations schematically in panel (k) in the lower part of Fig. \ref{fig:four}. 

TaIrTe$_4$ thus emerges as a very interesting, from both fundamental and practical points of view, material. Although the Weyl points themselves lie above the Fermi level, the topological surface states which support the Fermi surface arcs are clearly present and reveal a unique spin-texture. Unlike the surface states of topological insulators the spins of which are directed tangential to the closed FS contour, the spin-texture of arcs in TaIrTe$_4$ is strongly unidirectional implying a very anisotropic spin transport properties and even a possibility of storing the information. Indeed, both surfaces are characterized by the states with opposite spin separated in momentum space. Taking into account the strong anisotropy of the chiral anomaly expected in type-II Weyl semimetal, the experiments in magnetic and electric fields are urgently called for.

We thus demonstrated, that the predicted electronic structure of the non-centrosymmetric type-II Weyl semimetal TaIrTe$_4$ is realized experimentally. We have directly observed topological surface states and Fermi arcs with a unique spin texture, implying enormous application potential of the material.

This work was supported under DFG grant 1912/7-1. The authors acknowledge Diamond Light Source for the beamtime at I05 beamline under proposal SI13856 as well as the HiSOR Hiroshima for the beamtime at BL-9B beamline under proposal 16AG054.

\nocite{*}
  \printbibliography

\end{document}